\documentclass[11pt,a4paper]{article}
\pdfoutput=1
\usepackage{jheppub}
\usepackage[markup=nocolor]{changes}
\usepackage{hyperref}
\usepackage{url}
\usepackage[T1]{fontenc}

\usepackage{amssymb,amsbsy,epsfig,color,graphicx}
\usepackage{color}
\usepackage{[longtable}
\usepackage{array}
\usepackage{dcolumn}   
\usepackage{cellspace}
\usepackage{mathtools}
\usepackage{amstext}
\usepackage{amssymb}
\usepackage{stmaryrd}
\usepackage{stackrel}
\usepackage{graphicx}
\usepackage{esint}
\usepackage[utf8]{inputenc}
\usepackage{blindtext}
\usepackage{float}
\restylefloat{table}
\usepackage{booktabs}
\usepackage{enumitem} 
\usepackage{bbold}

\usepackage{etoolbox} 
\usepackage{lipsum} 
\usepackage[capitalize]{cleveref}
\usepackage{multirow}
\usepackage[caption=false]{subfig}
\renewcommand\[{\begin{equation}}
\renewcommand\]{\end{equation}}

\newcommand{\ba}{\begin{eqnarray}}
\newcommand{\ea}{\end{eqnarray}}

\newcommand{\mfund}{{\cal M}}

\makeatletter

\appto{\appendix}{%
\@ifstar{\def\thesection{\unskip}\def\theequation@prefix{A.}}%
{\def\thesection{\Alph {section}}}%
}
\makeatother

\hypersetup{pdfstartview=FitV,colorlinks=true,linkcolor=blue,citecolor=blue,filecolor=black,urlcolor=blue}

\title{\boldmath Nonlocal gravity with worldline inversion symmetry}
\author[a]{Steven Abel}
\author[c]{, Luca Buoninfante}
\author[d]{and Anupam Mazumdar}

\affiliation[a]{Institute for Particle Physics Phenomenology, Durham University, South Road, Durham, U.K.}
\affiliation[c]{INFN - Sezione di Napoli, Gruppo collegato di Salerno, I-84084 Fisciano (SA), Italy}
\affiliation[d]{Van Swinderen Institute, University of Groningen, 9747 AG, Groningen, The Netherlands}

\abstract{We construct a quadratic curvature theory of gravity whose graviton propagator around the Minkowski background respects wordline inversion symmetry, the particle approximation to modular invariance in string theory. This symmetry automatically yields a corresponding gravitational theory that is nonlocal, with the action containing infinite order differential operators. As a consequence, despite being a higher order derivative theory, it is ghost-free and has no degrees of freedom besides the massless spin-$2$ graviton of Einstein's general relativity. By working in the linearised regime we show that the point-like singularities that afflict the (local) Einstein's theory are smeared out.
}

\keywords{}

\begin{document}

\maketitle


\section{Introduction}

Einstein's general relativity (GR) is the most widely studied theory of gravity, and its predictions have been tested to very high precision in the infrared (IR) regime, i.e. at large distances and late times \cite{-C.-M.}. Despite passing these tests, there are unsolved conceptual problems which indicate that Einstein's GR is merely an effective field theory of gravitation: it works very well at low energy but breaks down in the ultraviolet (UV). Indeed at the classical level the Einstein-Hilbert Lagrangian, $\sqrt{-g}\mathcal{R},$ suffers from the presence of blackhole and cosmological singularities \cite{Hawking} (implying problems in the short-distance regime), while at the quantum level it is non-renormalisable from a perturbative point of view (implying problems in the high-energy regime) \cite{tHooft:1974toh,Goroff:1985th}. Therefore there is a consensus that ultimately GR will need to be extended.

One possible extension of GR is to add terms that are quadratic in curvature, such as $\mathcal{R}^2$ and $\mathcal{R}_{\mu\nu}\mathcal{R}^{\mu\nu}.$ The resulting actions are power counting renormalisable as shown in Ref.~\cite{-K.-S.}. However they are still non-physical because of the presence of a massive spin-$2$ ghost degree of freedom which classically  causes Hamiltonian instabilities, and which quantum mechanically breaks the unitarity condition of the S-matrix.

The appearance of ghost modes is related to the presence of higher order time derivatives in the field equations \cite{Ostrogradsky:1850fid}. However it is known that these unwelcome degrees of freedom can be avoided in higher derivative theories if the order of the derivatives is not finite but {\it infinite}. By introducing certain non-polynomial differential operators into the action, for example $e^{\Box/\mfund^2}$ with $\mfund$ being a new fundamental scale, one can prevent the appearance of extra poles in the physical spectrum \cite{Krasnikov,Kuzmin,Moffat,Tomboulis:1997gg}, because the presence of non-polynomial derivatives makes the action {\it nonlocal}. In fact such nonlocal models were the subject of very early studies, in which it was noted that they can improve the UV behavior of loop integrals (see Refs. \cite{efimov}). 

This promising property motivated deeper exploration of these nonlocal or so-called  {\it infinite derivative} field theories. 
The first relevant applications in the gravitational context were made in Refs. \cite{Biswas:2005qr,Modesto:2011kw,Biswas:2011ar,Biswas:2016etb} which demonstrated the possibility of constructing a quadratic curvature theory of gravity that is classically stable and unitary at the quantum level. It has also been noted that nonlocality can regularise infinities, and many efforts have been made to resolve black hole \cite{Modesto:2011kw,Biswas:2011ar,Biswas:2013cha,Edholm:2016hbt,Frolov:2015bia,Frolov,Frolov:2015usa,Buoninfante:2018xiw,Koshelev:2018hpt,Buoninfante:2018rlq,Buoninfante:2018stt,Buoninfante:2018xif,Buoninfante:2019swn,Kilicarslan:2018yxd,Kilicarslan:2019njc} and cosmological \cite{Biswas:2005qr,Biswas:2010zk,Biswas:2012bp,Koshelev:2012qn} singularities. Furthermore, renormalisability \cite{Modesto:2014lga,Talaganis:2014ida,Ghoshal:2017egr}, causality \cite{Tomboulis:2015gfa,Buoninfante:2018mre}, unitarity \cite{sen-epsilon,carone,Briscese:2018oyx,chin,Christodoulou:2018jbn}, scattering amplitudes \cite{Biswas:2014yia,Dona:2015tra,Buoninfante:2018gce}, spontaneous breaking of symmetry \cite{Gama:2018cda,Hashi:2018kag} and counting of initial conditions \cite{Barnaby:2007ve,Calcagni:2018lyd} have also been discussed and analysed. Further applications appear in the context of inflation \cite{inflation}, thermal field theory \cite{Biswas:2009nx,Biswas:2010xq,Biswas:2010yx} and Galilean theories \cite{Buoninfante:2018lnh}.

In the present work we are motivated by this kind of nonlocal field theory as an approximation to string theory. Indeed nonlocal theories have made an appearance in the context of both string field theory \cite{Witten:1985cc,eliezer,Tseytlin:1995uq,Siegel:2003vt} and p-adic strings \cite{Freund:1987kt}. Focussing on this particular aspect, one may first ask what is the best language in which to formulate the nonlocal field theory  approximation to a particular string theory? In \cite{Abel:2019ufz,AbelLewis} it was suggested that one should most naturally be working within the worldline formalism \cite{Feynman:1950ir,Strassler:1992zr,Schmidt:1994zj,Schubert:2001he,shubert}. Indeed the immediate outcome of discarding the higher modes of a first-quantised string theory in order to get a particle approximation is precisely a worldline theory with corrections that render it nonlocal at the fundamental scale \cite{AbelLewis}.

What particular nonlocal field theories might be legitimate particle approximations to string theory?
There are actually two factors in a string amplitude that can be responsible for its good UV behaviour and that one might wish to imitate in a nonlocal theory: the partition function, and the world-sheet Green's function. Which one is dominant  depends on the kinematics. The partition function governs the regularisation when the external momenta are low but the corresponding particle diagram would be UV divergent. For example the effective potential of non-supersymmetric (but non-tachyonic) string theories is necessarily rendered finite by the partition function. Such regularisation has been mimicked in particle theories by so-called ``minimal length'' theories \cite{Padmanabhan:1996ap}.  On the other hand when the external momenta (or rather their kinematic invariants) are very large compared to the string scale it is the world-sheet Green's functions that 
soften the amplitudes. Their short distance behaviour is known  exponentially to   suppress string amplitudes even at tree level  \cite{Gross:1987kza}. 

In the case of closed strings at loop-order, {\it modular invariance} can be identified as the key element that is operating in both cases. This symmetry governs both the partition function and the Green's function. In the case of the former, modular invariance induces duality symmetries in the space-time, whose effect can be modelled by making the aforementioned ``minimal-length''   adjustment to the partition function \cite{Padmanabhan:1996ap}. Here we will  instead focus on modelling the softening behaviour of the Green's function.

In Ref.~\cite{Abel:2019ufz} it was suggested that theories with worldline inversion symmetry are the best way to mimick the regularising properties
 of modular invariance. To see why, 
we can begin by considering a generic two-point one-loop integral in string theory. In the  ``particle limit'' (i.e. $\tau_2\gg 1$) it will collapse to the following heuristic form (see \cite{AbelLewis}):
\begin{equation}
{\cal A} \sim \int_0^1 dx dy \int_0^1 d\tau_1 \int_{\sim 1}^\infty \frac{d \tau_2}{\tau_2^2}   {\cal Z(\tau) } \,e^{- s x (1-x) \pi \alpha' \tau_2 +\ldots }~,
\end{equation}
where $\tau=\tau_1+i \tau_2$ is the modular parameter, and $z=x+i y$ is the displacement between the two vertices. In the above the 
exponent is what is left of the Green's function at large $\tau_2$, while $s = k_1\cdot k_2$ is the kinematic invariant\footnote{Having the correct conformal weight for the vertices requires $k_1^2=k_2^2=0$ and being on shell would imply $s=0$, so we are implicitly employing the usual trick of slightly violating Lorentz invariance to retain explicit dependence on $s$.}. We will ignore the accompanying pre-factors because these would be the same as in the effective particle theory. In this large $\tau_2$ region of the fundamental domain the $y$ and $\tau_1$ integrals become ``inert'' with the latter simply enforcing the level-matching of the physical spectrum. The whole integral is then ``projected'' to a worldline integral over $x$ (the usual Feynman parameter) and $\pi \alpha' \tau_2\equiv t$ (the usual Schwinger parameter).  In other words 
the particle limit yields a result directly in the worldline formalism,
\begin{equation}
\label{a1}
{\cal A} ~\sim \sum_{i = \rm  \scriptscriptstyle  physical} \int_0^1 dx \int_{\sim \alpha'}^\infty \frac{d t}{t}  \,e^{- ( s x (1-x) + m_{i}^2) t +\ldots }~,
\end{equation}
where the $m_i^2$ term in the exponent drops out of the partition function ${\cal Z}(\tau)$ and where $\sqrt{\alpha'}$ is the string-length.

This encapsulates the effective particle theory contribution to the string amplitude. But note that invariance of the whole amplitude under the  $\tau \rightarrow -1/\tau$ modular transformation 
means that one could equally write the integral in the domain where it approaches the cusp at $\tau \rightarrow 0$:
\begin{equation}
\label{a2}
{\cal A} ~\sim \sum_{i = \rm  \scriptscriptstyle  physical} \int_0^1 dy \int_0^{\sim \alpha'}  \frac{d t}{t}   \,e^{- ( s y (1-y) + m_{i}^2) \frac{1}{{\mfund}^4 t} +\ldots }~,
\end{equation}
where we define $\mfund ^2 =1/\pi \alpha' $. In this limit it is $y$ rather than $x$ that drops out of the worldsheet Green's function to end up playing the role of the Feynman parameter. But since 
the other variable is inert, the integral as $t\rightarrow 0$ ~---\,\,which is  a copy of \eqref{a1}\,\,---~ can just as well be interpreted as  continuing the $t$-integral into the deep UV, but with $t\rightarrow 1/{(\mfund^4 t)}$. In \cite{Abel:2019ufz} this was used to argued that one can capture the behaviour of the entire amplitude by writing a nonlocal theory with $t$ replaced by $T(t) = t+1/\mfund^4 t$ and integrating over {\it all} $t$. This approximation reproduces the 
asymptotic behaviour at the IR and UV cusps. It is reminiscent of string theory in the sense that the deep UV is identified as just another IR, with the 
difference being that in  the full string theory there are an infinite number of fundamental domains not just two.  
In summary the gross UV/IR mixing behaviour of strings (and modular invariance)  can be mimicked in the particle context by suitably modifying the Klein-Gordon propagator so that it exhibits a worldline {\it inversion symmetry}, $t\rightarrow 1/(\mfund^4 t)\,.$ 

In this paper we will explore such string theory inspired nonlocal field theories in the gravitational context. Our aim is to formulate a gravitational theory whose propagator around Minkowski space exhibits the above worldline inversion symmetry, and 
to investigate some of its consquences. This is possible despite the technicalities of writing the higher spin components of the theory in the worldline formalism \cite{Schubert:2001he}. The only price to pay is the introduction of nonlocality.

The paper is organized as follows. In Section \ref{scalar-modular-invariance} we briefly review how to introduce the worldline inversion symmetry at the level of the propagator for a simple scalar field, and also show its regularising properties. In Section \ref{new-gravity} we formulate a theory of gravity with worldline inversion symmetry built in and show that such an imposition automatically requires that the Lagrangian has to be infinite order in derivatives (i.e. nonlocal) but still ghost-free. In Section \ref{grav-pot}, we solve the linearised field equations in the presence of a static point-like source and explicitly show how spacetime singularities can be avoided due to the presence of nonlocality. Finally, in Section \ref{conclus} we draw our conclusions and discuss the outlook.

Throughout the paper we adopt the mostly positive convention for metric signature, $(-+++),$ and we work in Natural Units $\hbar=1=c.$


\section{Scalar propagator and worldline inversion symmetry}\label{scalar-modular-invariance}

We begin in this section by recapping and extending the scalar field Euclidean propagator, which can be defined in a general way in momentum space as an integral over a single real worldline parameter $t$ as follows \cite{Abel:2019ufz}:
\begin{equation}
\Pi(p^2)~=~\int_{0}^{\infty} dt e^{-T(t)(p^2+m^2)}~.\label{propag}
\end{equation}
The {\it proper-time} function $T(t)$ uniquely defines the propagator in momentum space. We can immediately see that  $T(t)=t$ gives the standard Schwinger parametrization for the Klein-Gordon propagator $1/(p^2+m^2)\,.$ 

The parameter $t$ has dimensions of length-squared, and therefore modified propagators can only be characterised by a non-trivial $T(t)$ at the expense of adding a new fundamental scale. For instance, the exponential propagator that appears in string theory \cite{Tseytlin:1995uq,Siegel:2003vt} and infinite derivative theories \cite{Biswas:2005qr,Tomboulis:1997gg,Biswas:2011ar,Modesto:2011kw} can be recovered as
\begin{equation}
T(t)~=~t+\frac{1}{\mfund^2}\quad ~\Rightarrow ~\quad \Pi(p^2)~=~\frac{e^{-(p^2+m^2)/\mfund^2}}{p^2+m^2}~.\label{exp-propag}
\end{equation}
Note that any modification must have a consistent IR limit, which means that the proper-time must satisfy
\begin{equation}
\lim\limits_{t\rightarrow \infty} \frac{T(t)}{t}~=~1\,.\label{IR limit}
\end{equation}
In addition we require the propagator to be ghost-free, i.e. we require that no negative norm states are present. A sufficient condition for this was found in Ref.\cite{Abel:2019ufz}:
\begin{equation}
{\rm Re}\left\lbrace T(t)\right\rbrace>0\quad\forall t>0 \quad {\rm and} \quad  tT(t^{-1})\quad{\rm is}\,\,\,{\rm entire}\,.\label{ghost-freeness}
\end{equation}
Following  Ref.\cite{Abel:2019ufz} and the introduction, we can mimick 
the inversion M\"obius transformation of the modular group by imposing inversion symmetry at the level of the proper-time function:
\begin{equation}
t~\rightarrow ~\frac{1}{\mfund^4 t}~,\label{inv-symm}
\end{equation}
where  ${\cal M}$ is the fundamental scale required for dimensionality. 
It is straightforward to prove that the only proper-time function satisfying  both \eqref{IR limit} and the ghost-freeness condition \eqref{ghost-freeness}, that is also invariant under \eqref{inv-symm}, is
\begin{equation}
T(t)~=~ t+\frac{1}{{\cal M}^{\prime\,2 } } + \frac{1}{\mfund^4 t}~,\label{only-possib}
\end{equation}
where ${\cal M}^{\prime}$ is a second constant parameter with dimensions of mass. We henceforth set ${\cal M}^{\prime\, }\rightarrow \infty \,,$ so that all information on new physics is encapsulated in a single fundamental scale $\mfund$. 

We are now able to compute the corresponding the scalar propagator by plugging the expression \eqref{only-possib} (with $1/\mfund^{\prime 2}  =0$) into  \eqref{propag} to find  \cite{Abel:2019ufz}:
\begin{equation}
\begin{array}{rl}
\Pi(p^2)~=~&\displaystyle \frac{2}{\mfund^2}{\rm K}_1\left(\frac{2(p^2+m^2)}{\mfund^2}\right)~\equiv~ \frac{1}{f(p^2)}\frac{1}{p^2+m^2}~,\\[3mm] 
f(p^2)~\equiv~ &\displaystyle \frac{\mfund^2}{2(p^2+m^2){\rm K}_1\left(2(p^2+m^2)/\mfund^2\right)}~,\label{propag-invar}
\end{array}
\end{equation}
where ${\rm K}_1\left(z\right)$ is the modified Bessel function of the second kind.
At low energy, $p^2/\mfund^2\ll 1,$ the propagator \eqref{propag} tends to the correct IR limit, i.e. $1/(p^2+m^2)$, while in the high energy regime, $p^2/\mfund^2\gg 1,$ it shows an exponentially suppressed behaviour (recalling that we are in Euclidean space):
\begin{equation}
\Pi(p^2)~\xrightarrow{\rm UV} ~\frac{\sqrt{\pi}e^{-2(p^2+m^2)/\mfund^2}}{\sqrt{p^2+m^2}} ~.\label{uv-regime}
\end{equation}
Note that the UV behaviour of the amplitudes is regularised through exponential suppression, in accord with the usual string picture.

Moreover, by analyzing the Bessel function in \eqref{propag-invar} we see that no extra pole is present besides the standard one at $p^2=-m^2,$ so the propagator is ghost-free. Indeed, ${\rm K}_1(z)$ is a holomorphic function in the right-half complex plane and has a branch cut for ${\rm Re}\left\lbrace z\right \rbrace <0$ starting at $z=0\,$, in agreement with the structure deduced in \cite{Abel:2019ufz}.

As expected the function $f(p^2)$ in \eqref{propag-invar} is non-polynomial in the momentum $p^2,$ implying a non-polynomial differential operator in coordinate space. In fact, one can define the corresponding nonlocal action for a scalar field as
\begin{equation}
S~=~\frac{1}{2}\int d^4x\, \phi(x)\,\Pi^{-1}\left(-\Box\right)\phi(x)~,\label{lagrang}
\end{equation}
where the operator
\begin{equation}
\Pi(-\Box)~=~\frac{2}{\mfund^2}{\rm K}_1\left(\frac{2(-\Box+m^2)}{\mfund^2}\right)~,\label{propag-coord-space}
\end{equation}
is made up of infinite order derivatives.

\subsection{Propagator in coordinate space}

As a warm-up for the gravitational case, it is useful  now to obtain the Euclidean propagator in coordinate space, and study its short-distance behaviour. This is defined as
\begin{equation}
\begin{array}{rl}
\Pi(x)~~=~ & \displaystyle \int \frac{d^4 p}{(2\pi)^4}\Pi(p)e^{ip\cdot x}\\[3mm]
~=~& \displaystyle \frac{2}{\mfund^2}\int \frac{d^4 p}{(2\pi)^4}{\rm K}_1\left(\frac{2(p^2+m^2)}{\mfund^2}\right)e^{ip\cdot x}~.
\end{array}
\end{equation}
Using polar coordinates in four dimensions, with $x = \sqrt{x^\mu x_\mu}$, 
we can recast the integral as
\begin{equation}
\Pi(x)~=~\frac{1}{2\pi^2\mfund^2x}\int_0^{\infty}dp\,p^2\,{\rm K}_1\left(\frac{2(p^2+m^2)}{\mfund^2}\right){\rm J}_1\left(px\right)~,\label{coord-integr}
\end{equation}
where ${\rm J}_1(x)$ is the ordinary Bessel function. The integral \eqref{coord-integr} cannot be performed analytically for $m\neq 0,$ but it can in the massless case, which yields 
\begin{equation}
\Pi(x)~=~\frac{\mfund^2}{64\pi}\left[{\rm I}_0\left(\frac{\mfund^2x^2}{8}\right)-{\rm L}_0\left(\frac{\mfund^2x^2}{8}\right)\right]~,
\end{equation}
where ${\rm I}_0(x)$ is a modified Bessel function of the first kind and ${\rm L}_0(x)$ is the modified Struve function. In the large distance regime, $\mfund x\gg 1,$ we recover the propagator\footnote{In taking the asymptotic limits of the Bessel and Struve functions one encounters Stoke's phenomenon, according to which subleading contributions are discontinuous in certain regions of the complex plane. However, this does not disrupt the consistency of the IR limit which can be checked graphically.} of the normal local theory, $1/(4\pi^2x^2)$, while in the short-distance regime, $\mfund x\ll 1,$ the propagator is regularised as 
\begin{equation}
\lim\limits_{x\rightarrow 0}\Pi(x)~=~\frac{\mfund^2}{64\pi}~.
\end{equation}
As anticipated the $1/x^2$ singularity of the standard local field theory is smoothed out and regulated by the nonlocality. The same phenomenon can be observed  in the massive case, by computing  the $m\neq 0$ integral in Eq.\eqref{coord-integr} numerically.


\section{Nonlocal gravitational theory}\label{new-gravity}

Let us now construct an analogous  gravitational theory whose propagator around the Minkowski background exhibits the same worldline inversion symmetry. In other words, we will identify the gravitational action whose linearised version gives a modified graviton propagator that has a similar structure to that in Eq.\eqref{propag-invar} for a scalar field. 

Since we aim to work with an action containing terms quadratic in the curvature tensors, let us first introduce some fundamental tools. The most general parity-invariant and torsion-free quadratic curvature action around a maximally symmetric background and up to second order variation in the metric perturbation is given by~\cite{Biswas:2011ar,Biswas:2016etb}:
\begin{equation}
S~=~\displaystyle \!\frac{1}{2\kappa^2}\int d^4x\sqrt{-g}\left\lbrace \mathcal{R}+\frac{1}{2}\left[\mathcal{R}F_1(\Box)\mathcal{R}+\mathcal{R}_{\mu\nu}F_2(\Box)\mathcal{R}^{\mu\nu}+\mathcal{R}_{\mu\nu\rho\sigma}F_3(\Box)\mathcal{R}^{\mu\nu\rho\sigma}\right]\right\rbrace ~,
\label{quad-action}
\end{equation}
where $\kappa:=\sqrt{8\pi G},$ with $G=1/M_p^2$ being the Newton constant, and $F_i(\Box)$ being two differential operators which can be uniquely determined around the Minkowski background by fixing the form of the graviton propagator~\cite{Modesto:2011kw,Biswas:2011ar}. 

Note that, as we are interested in second order metric perturbations around the Minkowksi metric, we are always allowed to neglect the Riemann squared term $\mathcal{R}_{\mu\nu\rho\sigma}F_3(\Box)\mathcal{R}^{\mu\nu\rho\sigma}$ up to this order. Indeed, one can show that the following relation holds for any power $n$ of the d'Alembertian:
\begin{equation}
\mathcal{R}_{\mu\nu\rho\sigma}\Box^n\mathcal{R}^{\mu\nu\rho\sigma}~=~4\mathcal{R}_{\mu\nu}\Box^n\mathcal{R}^{\mu\nu}-\mathcal{R}\Box^n\mathcal{R}+\mathcal{O}(\mathcal{R}^3)~+~{\rm div}~,\nonumber
\end{equation}
where $\mathcal{O}(\mathcal{R}^3)$ stands for higher order contributions $\mathcal{O}(h^3)$ and {\rm div} stands for total derivatives. Thus, the gravitational action in Eq.\eqref{quad-action} can be written as
\begin{equation}
S~=~\displaystyle \!\frac{1}{2\kappa^2}\int d^4x\sqrt{-g}\left\lbrace \mathcal{R}+\frac{1}{2}\left[\mathcal{R}\mathcal{F}_1(\Box)\mathcal{R}+\mathcal{R}_{\mu\nu}\mathcal{F}_2(\Box)\mathcal{R}^{\mu\nu}\right]+\mathcal{O}\left(\mathcal{R}^3\right)\right\rbrace~ ,
\label{quad-action-reduced}
\end{equation}
where we have defined
\begin{equation}
\mathcal{F}_1(\Box)~=~F_1(\Box)-F_3(\Box)~,\quad\quad  \mathcal{F}_2(\Box)~=~F_2(\Box)+4F_3(\Box)~.
\end{equation}
By perturbing around the Minkowski metric,
\begin{equation}
g_{\mu\nu}~=~\eta_{\mu\nu}+\kappa h_{\mu\nu}~, \label{lin-metric}
\end{equation}
where  $h_{\mu\nu}$ is the metric perturbation, we obtain the following linearised gravitational action up to order $\mathcal{O}(h_{\mu\nu}^2)$ \cite{Biswas:2011ar}:
\begin{equation}
\begin{array}{rl}
S^{(2)}~=~&\displaystyle \frac{1}{4}\int d^4x\left\lbrace \frac{1}{2}h_{\mu\nu}f(\Box)\Box h^{\mu\nu}-h_{\mu}^{\sigma}f(\Box)\partial_{\sigma}\partial_{\nu}h^{\mu\nu}\right.\\[3mm]
&\displaystyle \,\,\,\,\,\,\,\,\,\,\qquad\qquad\qquad\qquad -~\frac{1}{2}hg(\Box)\Box h+hg(\Box)\partial_{\mu}\partial_{\nu}h^{\mu\nu}\\[3mm]
&\,\,\,\,\,\,\,\,\,\,\displaystyle\qquad\qquad\qquad\qquad\qquad\qquad\qquad\left.+~\frac{1}{2}h^{\lambda\sigma}\frac{f(\Box)-g(\Box)}{\Box}\partial_{\lambda}\partial_{\sigma}\partial_{\mu}\partial_{\nu}h^{\mu\nu}\right\rbrace\\[3mm]
~\equiv~&\displaystyle  \frac{1}{4}\int d^4x h_{\mu\nu}\mathcal{O}^{\mu\nu\rho\sigma}h_{\rho\sigma}~,
\label{lin-quad-action}
\end{array}
\end{equation}
with the kinetic operator defined as
\begin{equation}
\begin{array}{rl}
\mathcal{O}^{\mu\nu\rho\sigma}~\equiv~ &\displaystyle \frac{1}{4}\left(\eta^{\mu\rho}\eta^{\nu\sigma}+\eta^{\mu\sigma}\eta^{\nu\rho}\right)f(\Box)\Box-\frac{1}{2}\eta^{\mu\nu}\eta^{\rho\sigma}g(\Box)\Box \\[3mm]
&\displaystyle \qquad -~\frac{1}{4}\left(\eta^{\mu\rho}\partial^{\nu}\partial^{\sigma}+\eta^{\mu\sigma}\partial^{\nu}\partial^{\rho}
+\eta^{\nu\rho}\partial^{\mu}\partial^{\sigma}+\eta^{\nu\sigma}\partial^{\mu}\partial^{\rho}\right)f(\Box) \\[3mm]
&\displaystyle \qquad \qquad +~\frac{1}{2}\left(\eta^{\mu\nu}\partial^{\rho}\partial^{\sigma}+\eta^{\rho\sigma}\partial^{\mu}\partial^{\nu}\right)g(\Box)
~+~\frac{1}{2}\frac{f(\Box)-g(\Box)}{\Box}\partial^{\mu}\partial^{\nu}\partial^{\rho}\partial^{\sigma}~. 
\label{kinetic-grav}
\end{array}
\end{equation}
Here $h\equiv\eta_{\mu\nu}h^{\mu\nu}$ stands for the trace and $\Box=\eta^{\mu\nu}\partial_{\mu}\partial_{\nu}$ is the flat d'Alembertian operator, while the functions
\begin{equation}
f(\Box)~=~\displaystyle  1+\frac{1}{2}\mathcal{F}_2(\Box)\Box~,\quad
g(\Box)~=~ 1-2\mathcal{F}_1(\Box)\Box-\frac{1}{2}\mathcal{F}_2(\Box)\Box \label{relat-form-fact}~,
\end{equation}
are combinations of the two form factors $\mathcal{F}_i(\Box)\,.$

By inverting the kinetic operator in Eq.\eqref{kinetic-grav}, and after having introduced a suitable gauge fixing term, one can obtain the propagator around a Minkowski background whose saturated part is given by \cite{Krasnikov,Tomboulis:1997gg,Biswas:2011ar,Modesto:2011kw}:
\begin{equation}
\Pi_{\rm GR, \mu\nu\rho\sigma}(p) ~=~\frac{\mathcal{P}_{\mu\nu\rho\sigma}^2}{f(p)p^2}+\frac{\mathcal{P}_{s,\mu\nu\rho\sigma}^0}{(f(p)-3g(p))p^2}~,
\end{equation}
where the spin projection operators  $\mathcal{P}_{\mu\nu\rho\sigma}^2$ and $\mathcal{P}^0_{s, \mu\nu\rho\sigma}$ project along the spin-$2$ and spin-$0$ components respectively, and are defined as \cite{VanNieuwenhuizen:1973fi,Biswas:2013kla}
\begin{equation}
\begin{array}{ll}
\mathcal{P}^2_{\mu\nu\rho\sigma}~=~\displaystyle \frac{1}{2}\left(\theta_{\mu\rho}\theta_{\nu\sigma}+\theta_{\mu\sigma}\theta_{\nu\rho}\right)-\frac{1}{3}\theta_{\mu\nu}\theta_{\rho\sigma}~,\quad\quad \mathcal{P}^0_{s,\mu\nu\rho\sigma}~=~\displaystyle \frac{1}{3}\theta_{\mu\nu}\theta_{\rho\sigma}~,&\\[3mm]
\,\,\,\,\,\,\,\,\,\,\,\,\,\,\,\,\,\,\,\,\,\,\,\,\,\,\,\,\,\,\,\,\,\,\,\,\,\,\,\,\,\,\,\,\displaystyle \theta_{\mu\nu}~=~\eta_{\mu\nu}-\omega_{\mu\nu}~,\quad\displaystyle \quad \omega_{\mu\nu}~=~\frac{k_{\mu}k_{\nu}}{k^2}~.&
\label{spin-proj}
\end{array}
\end{equation}
As a consistency check, note that for $f=1=g$ we recover the saturated part of the Einstein-Hilbert propagator \cite{VanNieuwenhuizen:1973fi,Biswas:2013kla},
\begin{equation}
\Pi_{\rm GR, \mu\nu\rho\sigma}(p)~=~\frac{\mathcal{P}_{\mu\nu\rho\sigma}^2}{p^2}-\frac{\mathcal{P}^0_{s, \mu\nu\rho\sigma}}{2p^2}~.
\end{equation}
%

\subsection{Graviton propagator with worldline inversion symmetry}

In order to find the gravitational analogue of the propagator in Eq.\eqref{propag-invar} we work in the simplest case in which the only on-shell propagating degrees of freedom are the massless transverse graviton  of Einstein's general relativity with helicities $\pm 2$. This requirement corresponds to the condition 
\begin{equation}
f(\Box)~=~g(\Box)~,\quad\quad 2\mathcal{F}_1(\Box)~=~-\mathcal{F}_2(\Box)~,\label{only spin-2}
\end{equation}
which implies
\begin{equation}
\Pi_{\rm GR, \mu\nu\rho\sigma}(p)~=~\frac{1}{f(p^2)}\left(\frac{\mathcal{P}_{\mu\nu\rho\sigma}^2}{p^2}-\frac{\mathcal{P}^0_{s, \mu\nu\rho\sigma}}{2p^2}\right)~,
\end{equation}
whose functional form is similar to the one in \eqref{propag-invar}: indeed we have the standard local propagator multiplied by some function $1/f(p^2).$ Therefore, by choosing the function $f(p^2)$ as in \eqref{propag-invar} we obtain the following worldline inversion invariant graviton propagator:
\begin{equation}
\Pi_{\rm GR, \mu\nu\rho\sigma}(p)~=~\frac{2}{\mfund^2}{\rm K}_1\left(\frac{2p^2}{\mfund^2}\right)\left[\mathcal{P}_{\mu\nu\rho\sigma}^2-\frac{1}{2}\mathcal{P}^0_{s, \mu\nu\rho\sigma}\right]\,,\label{grav-prop}
\end{equation}
which again is ghost-free as it possesses only one single pole at $p^2=0$ corresponding to the usual massless graviton degree of freedom. {This procedure allows us to circumvent the technicalities of writing the higher spin components of the theory in the worldline formalism because the spin-$0$ part of the propagator necessarily governs the whole structure \cite{Schubert:2001he}\footnote{Typically higher spins would be represented on the worldline as a global supersymmetry relevant for each diagram, with an additional supersymmetry introduced  for each half-unit of spin.
 Even though the supersymmetry is in principle broken by the periodicity conditions of the 
diagram, the breaking can be considered to be a spontaneous one from a one-dimensional point of view (with the parameter T playing the 
role of a 1D compactification modulus if it is a one-loop diagram). Hence 
the tensor structure in front of any amplitude must be independent of this breaking, implying that back in the usual field theory formalism 
the spin-$0$ component of the graviton propagator then determines the structure for the higher spins.}.
In fact, by working in the usual Feynman gauge one can show that the graviton propagator reads:
\begin{equation}
\begin{array}{rl}
\Pi_{\rm GR, \mu\nu\rho\sigma}(p)~=~&\displaystyle \left(\eta_{\mu\rho}\eta_{\nu\sigma}+\eta_{\mu\sigma}\eta_{\nu\rho}-\eta_{\mu\nu}\eta_{\rho\sigma}\right) \frac{1}{\mfund^2}{\rm K}_1\left(\frac{2p^2}{\mfund^2}\right)\\[3.5mm]
~=~&\displaystyle \frac{1}{2}\left(\eta_{\mu\rho}\eta_{\nu\sigma}+\eta_{\mu\sigma}\eta_{\nu\rho}-\eta_{\mu\nu}\eta_{\rho\sigma}\right)\int_{0}^{\infty} dt\, e^{-\left(t+\frac{1}{t\mathcal{M}^4}\right)p^2}~,
\end{array}
\end{equation}
for which invariance under the wordline inversion \eqref{inv-symm} is now manifest; in the limit $\mathcal{M}\rightarrow \infty$ we recover the GR graviton propagator in the Feynman gauge, as expected.}

From the form of the propagator, and so of the function $f(\Box),$ by using the relations in Eqs.(\ref{relat-form-fact},\ref{only spin-2}) we obtain
\begin{equation}
\displaystyle\mathcal{F}_1(\Box)~=~-\frac{1}{2}\mathcal{F}_2(\Box)~=~\displaystyle \frac{1}{\Box}+\frac{\mfund^2}{2\, \Box^2\,  {\rm K}_1(-\Box/\mfund^2)}~,
\label{form-factors}
\end{equation}
which gives the following gravitational action up to quadratic curvature terms:
\begin{equation}
\begin{array}{rl}
S~\,=~ &\displaystyle  \frac{1}{2\kappa^2}\int d^4x\sqrt{-g}\left\lbrace \mathcal{R}- G_{\mu\nu}\frac{1}{\Box}\mathcal{R}^{\mu\nu}\right.\\[3mm]
&\displaystyle\,\,\,\,\,\,\,\,\,\,\,\,\,\,\,\,\,\,\,\,\,\,\,\,\,\, \qquad\qquad\qquad\qquad\left.- \frac{\mfund^2}{2}G_{\mu\nu}\frac{1}{\Box^2\,  {\rm K}_1(-\Box/\mfund^2)}\mathcal{R}^{\mu\nu}\right\rbrace ~,
\end{array}
\label{nonlocal-quad-action}
\end{equation}
where we have introduced the Einstein tensor $G_{\mu\nu}=\mathcal{R}_{\mu\nu}-1/2g_{\mu\nu}\mathcal{R}$. Hence, the gravitational theory described by the action in Eq.\eqref{nonlocal-quad-action} is a higher (infinite) derivative theory of gravity which is ghost-free around a Minkowski background, despite the presence of higher order time derivatives. Up to quadratic curvature terms this is the unique action with spin-$2$ graviton propagator which exhibits invariance under the worldline inversion \eqref{inv-symm}. 



\section{Nonsingular gravitational potential}\label{grav-pot}

In this Section we wish to examine a physical implication of the nonlocality introduced by requiring the graviton propagator to be invariant under the worldline inversion symmetry in \eqref{inv-symm}. Namely we will demonstrate that the classical linearised spacetime metric in the presence of a static point-like source for the gravitational action \eqref{nonlocal-quad-action} is smoothed out. 

To begin we compute the linearised field equation for the action \eqref{nonlocal-quad-action}, which is 
\begin{equation}
\begin{array}{ll}
\displaystyle \frac{\mfund^2}{2\Box{\rm K}_1(-\Box/\mfund^2)}\left(\Box h_{\mu\nu}-\partial_{\sigma}\partial_{\nu}h_{\mu}^{\sigma}-\partial_{\sigma}\partial_{\mu}h_{\nu}^{\sigma}\right.\qquad\qquad&\\[3mm]
\displaystyle \,\,\,\,\,\,\,\,\,\,\,\,\,\,\,\,\,\,\,\,\,\,\qquad\qquad\qquad\left.+\eta_{\mu\nu}\partial_{\rho}\partial_{\sigma}h^{\rho\sigma}+\partial_{\mu}\partial_{\nu}h-\eta_{\mu\nu}\Box h\right)~=~16\pi G T_{\mu\nu}~,&
\end{array}
\label{lin-field-eq}
\end{equation}
where $T_{\mu\nu}$ is the stress-energy tensor describing the matter sector. By working in the Newtonian conformal gauge we can express the metric in isotropic coordinates as follows: 
\begin{equation}
ds^2~=~-(1+2\Phi)dt^2+(1-2\Phi)(dr^2+r^2d\Omega^2)~,\label{isotr-metric}
\end{equation}
so that  $\kappa h_{00}=-2\Phi$, $\kappa h_{ij}=-2\Phi\delta_{ij}$, $\kappa h=2(\Phi-3\Psi),$ with $\Phi$ being the gravitational potential. Moreover, for a static point-like source the stress-energy tensor acquires a very simple form, and indeed the only non-vanishing component is the density part: $T_{\mu\nu}=m\delta_{\mu}^0\delta_{\nu}^{0}\delta^{(3)}(\vec{r}),$ where $m$ is the mass of the object. Simplification due to staticity and spherical symmetry reveals that the only unknown, $\Phi(r)$, satisfies the following modified Poisson equation:
\begin{equation}
\frac{\mfund^2}{2{\rm K}_1(2\nabla^2/\mfund^2)}\Phi(r)~=~4\pi Gm\delta^{(3)}(\vec{r})~.
\label{field-eq-pot}
\end{equation}
As a consistency check, note that the local limit, $\mathcal{M}\rightarrow \infty$, recovers the standard Poisson equation whose solution is the Newtonian potential $\Phi(r)=-Gm/r\,.$ 

The equation in Eq.\eqref{field-eq-pot} is differential and of infinite order, which makes its solution in coordinate space quite complicated. It can be solved by Fourier transforming to momentum space through Fourier method and then transforming back, which gives 
\begin{equation}
\begin{array}{rl}
\Phi(r)~=~&\displaystyle -\frac{4 Gm}{\pi \mfund^2}\frac{1}{r}\int_0^{\infty}dk k\,{\rm sin}(kr)\,{\rm K}_1\left(\frac{2k^2}{\mfund^2}\right)\\[3mm]
~=~&\displaystyle \frac{Gm\mfund^2\pi}{16\sqrt{2}}r\left[{\rm I}_{\frac{1}{4}}^2\left(\frac{\mfund^2r^2}{16}\right)-{\rm I}_{-\frac{1}{4}}^2\left(\frac{\mfund^2r^2}{16}\right)\right]~,
\end{array}
\label{fourier-pot}
\end{equation}
where ${\rm I}_{\frac{1}{4}}(x)$ and ${\rm I}_{-\frac{1}{4}}(x)$ are modified Bessel functions of the first kind. By using standard relations 
\begin{figure}[t]
	\centering
	\includegraphics[scale=0.41]{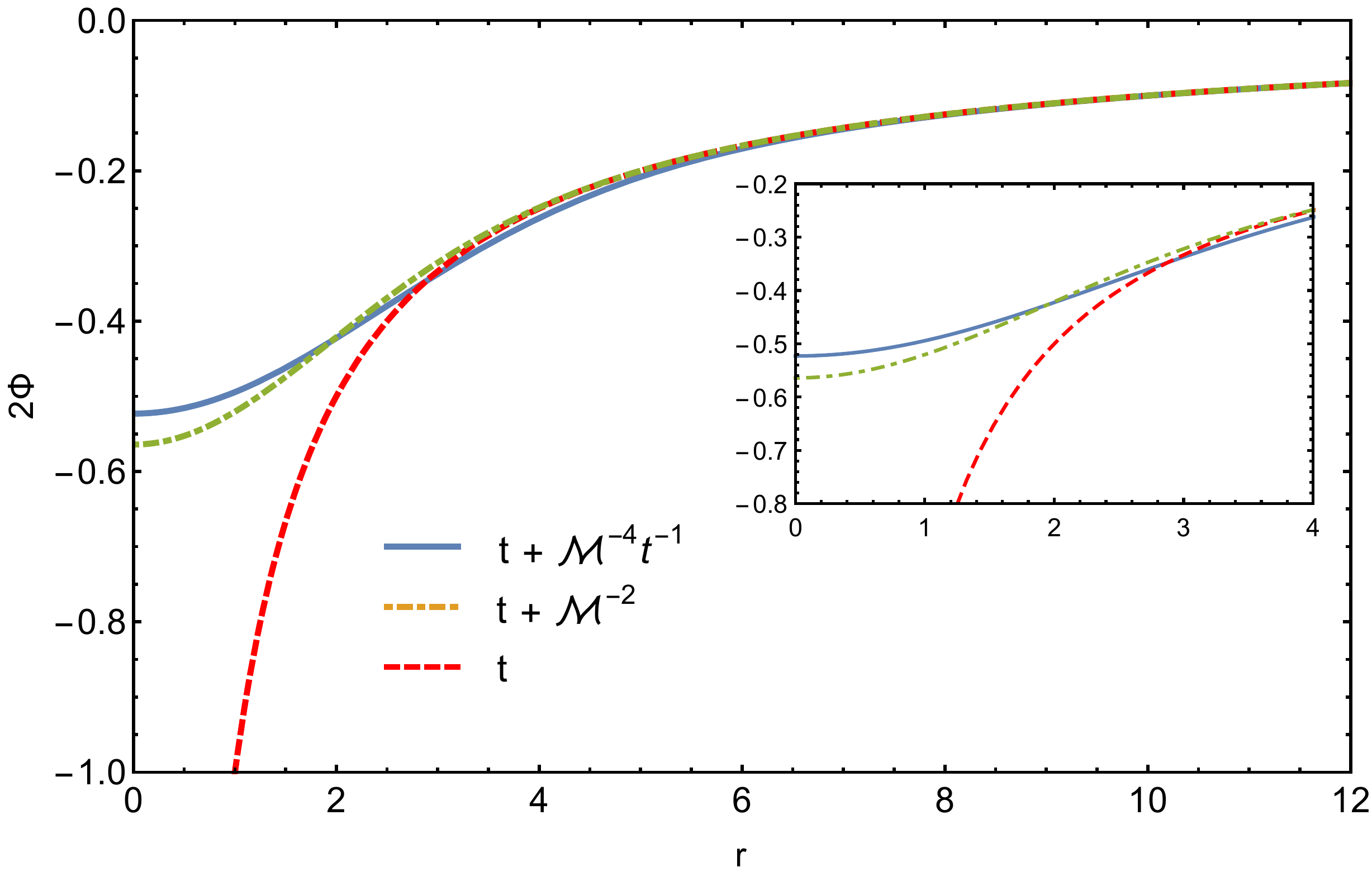}
	\protect\caption{The nonlocal gravitational potential (blue solid line, $t+\mathcal{M}^{-4}t^{-1}$) generated by a point-like source in the nonlocal theory described by the action \eqref{nonlocal-quad-action} and corresponding to the choice \eqref{only-possib} with $\mathcal{M}'=\infty$, compared to the Newtonian potential (red dashed line, $t$) and to the nonlocal potential corresponding to the choice \eqref{exp-propag} of the proper-time function (orange dotted-dashed line, $t+\mathcal{M}^{-2}$). Here we set $G=1=\mathcal{M}$ and $m=0.5$.}\label{fig1}
\end{figure}
%
%
the gravitational potential can also be recast as
\begin{equation}
\Phi(r)~=~ -\displaystyle \frac{Gm\mfund^2}{16}r \,{\rm K}_{\frac{1}{4}}\left(\frac{\mfund^2r^2}{16}\right)
%
\displaystyle \left[{\rm I}_{\frac{1}{4}}\left(\frac{\mfund^2r^2}{16}\right)+{\rm I}_{-\frac{1}{4}}\left(\frac{\mfund^2r^2}{16}\right)\right]\,.
\label{potential}
\end{equation}
%
At large distances $\mfund r\gg 1$ we recover the Newtonian behaviour as expected; while at short distances $\mfund r\ll 1$ the gravitational potential is regularised, and is entirely nonsingular at the origin:
\begin{equation}
\lim\limits_{r\rightarrow 0}\Phi(r)~=~-\frac{4 \pi Gm\mfund}{\Gamma^2(-1/4)}~,
\label{pot-r=0}
\end{equation}
where $\Gamma(x)$ is the Euler gamma function. We conclude that, as in Section \ref{scalar-modular-invariance} for the singularity-free scalar, the presence of nonlocality can be instrumental in resolving the gravitational singularities that afflict standard local theories. 

It is worth mentioning that, as the metric potential is monotonic with minimum at $r=0,$ the linear approximation can hold true from $r=0$ all the way up to $r=\infty$ provided the following inequality is satisfied:
\begin{equation}
2|\Phi|<1 \quad \Leftrightarrow \quad \frac{8\pi Gm\mfund }{\Gamma^2(-1/4)}< 1\,.
\end{equation}
{In Fig.\ref{fig1} we plot the gravitational potential \eqref{potential} along with Newton's potential (which is singular) and with the potential  corresponding to the choice \eqref{exp-propag} of proper-time function, which gives $\Phi_{(t+1/{\mathcal{M}^2})}(r)=-Gm{\rm Erf}(r\mathcal{M}/2)/r$ \cite{Biswas:2005qr,Biswas:2011ar,Buoninfante:2018xiw}. Note that both nonlocal potentials are strictly monotonic and are regularised at the origin with the only difference being that, in the case of worldline inversion symmetry, the spacetime metric describes a less compact gravitational system, namely $\Phi(0)<\Phi_{(t+1/{\mathcal{M}^2})}(0)$.}

{One can also check that all curvature invariants are non-singular at $r=0,$ so that no curvature singularities appear at all. For instance, the Kretschmann invariant $\mathcal{R}_{\mu\nu\rho\sigma}\mathcal{R}^{\mu\nu\rho\sigma}$ for the metric in Eq.(\ref{isotr-metric},\ref{potential}) is also finite at the origin, contrasting with the GR case in which it diverges as $1/r^6\,.$ Due to its lengthy expression we do not show it, but it is straightforward to show that the Kretschmann invariant tends to the following finite value in the short-distance regime:
\begin{equation}
\lim\limits_{r\rightarrow 0} \mathcal{R}_{\mu\nu\rho\sigma}\mathcal{R}^{\mu\nu\rho\sigma}\sim G^2m^2\mathcal{M}^6\,.
\end{equation}
}


\section{Conclusions}\label{conclus}

In this paper we have formulated a gravitational theory whose graviton propagator around the Minkowski background exhibits worldline inversion symmetry, which is a nonlocal particle mimicking the modular invariance of string theory. We showed that it is possible to construct such a propagator by following a quite straightforward procedure that circumvents the difficulties of writing the higher spin components of the theory in the worldline formalism \cite{Schubert:2001he}; see Eq.\eqref{grav-prop}. The price to pay is the introduction of non-polynomial differential operators in the gravitational action which necessarily becomes nonlocal, as shown in Eq.\eqref{nonlocal-quad-action}.

Despite there being higher derivatives, such a nonlocal theory of gravity is ghost-free at tree level, and indeed the propagator possesses only one pole at $p^2=0,$ and therefore no unhealthy degrees of freedom are present. Indeed the presence of infinite order derivatives ameliorates the short-distance behaviour of the theory entirely, and the modified gravitational potential turns out to be regularised at the origin in contrast to the Newtonian one which diverges in the limit $r\rightarrow 0$, very similar to the results in ghost free infinite derivative theories of gravity \cite{Biswas:2011ar}.


\subsection*{Acknowledgements} SAA would like to thank Nicola Dondi and Daniel Lewis for collaboration and discussion, and acknowledges support from the STFC and Royal-Society/CNRS International Cost Share Award IE160590. LB is grateful to Steve Abel for his warm hospitality at IPPP, Durham University, where this work was started. AM's research is supported by Netherlands Organization for Scientific Research (NWO) grant no. 680-91-119.



\end{document}